\documentclass[twocolumn,showpacs,preprintnumbers,amsmath,amssymb]{revtex4}

\usepackage{graphicx}
\usepackage{dcolumn}
\usepackage{bm}

\begin{document}

\preprint{APS/123-QED}

\title{Friction and the oscillatory motion of granular flows}

\author{L. Staron}

\affiliation{%
CNRS - Universit\'e Pierre et Marie Curie Paris 6, UMR 7190, Institut Jean Le Rond d'Alembert, F-75005 Paris, France.
}%


\begin{abstract}
This contribution reports on numerical simulations of 2D granular flows on erodible beds. The broad aim is to investigate whether simple flows of model granular matter exhibits spontaneous oscillatory motion in generic flow conditions, and in this case,  whether the frictional properties of the contacts between grains may affect the existence or the characteristics of this oscillatory motion.  The analysis of different series of simulations show that the flow develops an oscillatory motion with a well-defined frequency which increases like the inverse of the velocity's square root. We show that the oscillation is essentially a surface phenomena. The amplitude of the oscillation is higher for lower volume fractions, and can thus be related to the flow velocity and  grains friction properties. The study of the influence of the periodic geometry of the simulation cell shows no  significant effect. These results are  discussed in relation to sonic sands. 
\end{abstract}

\maketitle

\section{Introduction}

Granular flows have long been the subject of sustained interest due to their surprising properties: the existence of an internal yield stress allowing them both solid-like and fluid-like behaviors and more generally their elusive rheology \cite{volfson03,lois05,dacruz05,brewster08,wyart09,pouliquen99,gdrmidi04,campbell05,jenkins06}, their ability to segregate according to grain size or mass \cite{schroeter06,taberlet06,linares07,gray11,rivas11},  their similarities with glassy systems \cite{behringer08,liu10,tighe10,dauchot11},  their non-local properties \cite{aranson08,nichol10,staron10,reddy11}, to cite only few examples among many. While most of the features listed above have been the subject of careful laboratory experiments, some peculiarities of granular systems express themselves best in nature: this is the case of musical (or sonic) sands, and most famously, of booming dunes \cite{lindsay76,andreotti04,douady06,andreotti09,dagois10,dagois10b,hunt10}. Booming dunes, when a surface flow is triggered at their flank, can emit a surprisingly loud sound which has proven a long-lasting challenge to experimentalist and theoretician alike.  In spite of numerous advances, including experimental investigation, modeling and field measurements, the mechanisms at play in the booming phenomena remain controversial (see the recent review by \cite{andreotti12} and reference therein). However, converging observations have emerged. It was observed that the granular surface flow alone can produce the sound, without the dune being a necessary ingredient. Then, both laboratory experiments and field measurement show that a minimum flow velocity is needed for sound to be emitted. Moreover, sonic sand grains need specific contact properties, ensured in nature by a mineral coating  reproduced in the lab by successive bath in complex salty solutions \cite{dagois10b}.  Based on these observations,  two questions arises: can a simple granular flow spontaneously exhibit oscillatory motion  with a well-defined frequency? If yes, how do frictional contact properties affect the oscillatory dynamics?  These two questions are the subject of the present contribution. While intermittent motion close to jamming transition was already observed and discussed in relation to sonic sands \cite{silbert05,mills09,richard12}, we place ourselves in the case of rapid flows,  corresponding to the observation of a minimum velocity threshold for sound emission.  Applying discrete numerical simulation in 2D \cite{jean92,radjai96}, we simulate model granular flows on erodible beds; we analyze the dynamics of the simulated flows while varying the grains frictional properties over a range difficult  to attain in laboratory experiments.  Doing so, we show the existence of a spontaneous oscillation developing rapidly  and exhibiting a well-defined frequency which increases like the inverse of the velocity's square root. The amplitude of the frequency is higher for lower volume fraction, and can thus be related to both flow velocity and grains friction properties.  Finally, we investigate the influence of the spatial periodicity of the simulations on the above results and observe no significant effect. These results are  discussed in relation to sonic sands. 

\begin{figure}
\begin{minipage}{1.\linewidth} 
\centerline{\includegraphics[angle = 0,width = \linewidth]{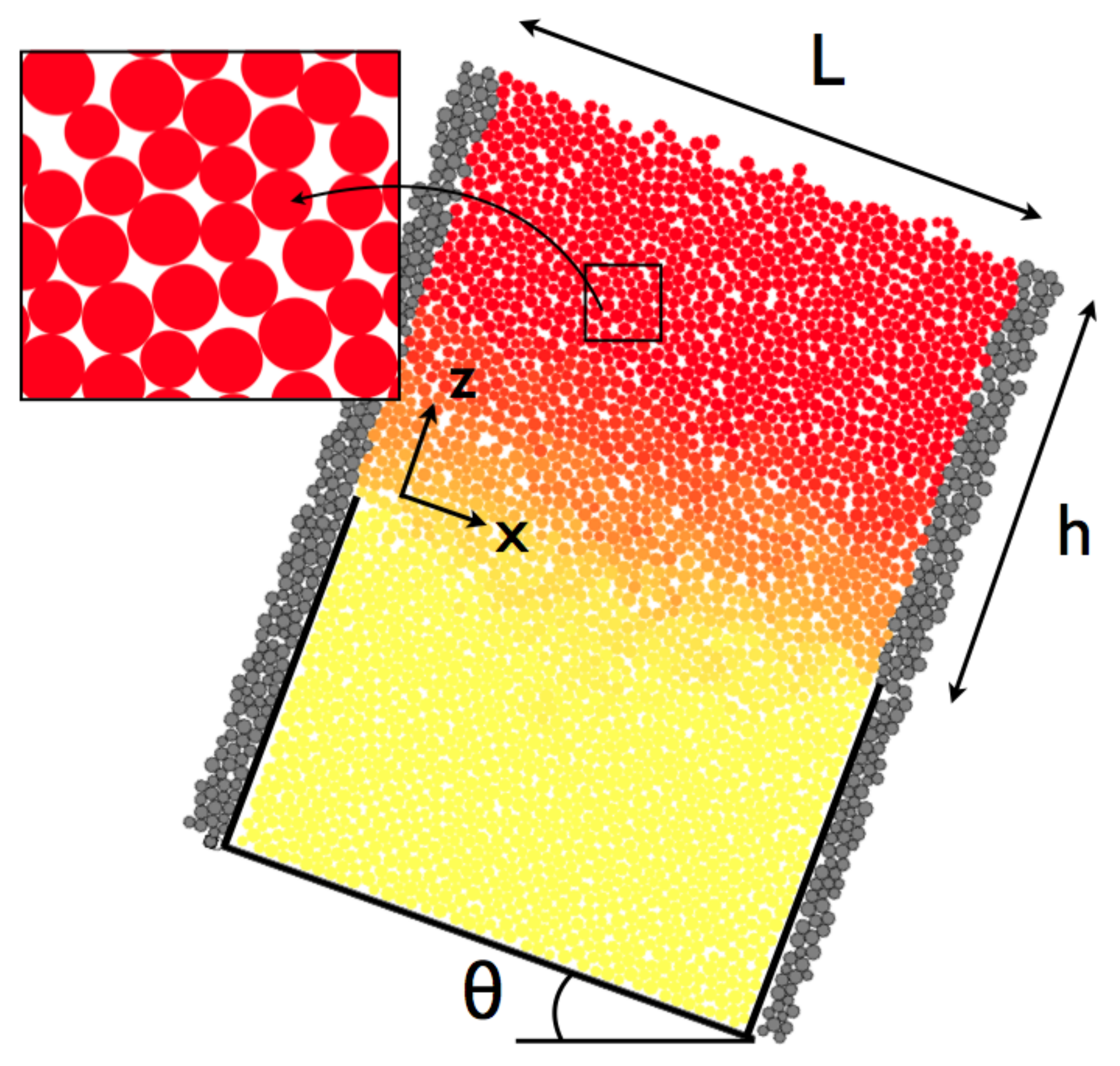}}
\end{minipage}
\caption{A 2D periodic flow on erodible bed tilted of an angle $\theta$ simulated by Contact Dynamics. The height of the flowing layer is denoted $h$ and the length of the sumulation cell is $L$. The yellow-red shade of the grains stands for their velocity (color on-line); the darker-grey shade shows images of the flowing grains through the periodic boundary condition.}
\label{Expe}
\end{figure}

\section{The Contact Dynamics simulations}

The simulations reported in this contribution were performed using the Contact Dynamics algorithm \cite{jean92,radjai09}. The grains are assumed to be perfectly rigid, which translates in a strict non-overlap condition at contact. They interact through a Coulombic friction law, relating the tangential force at each contact $f_t$ to the normal force at the same contact $f_t$ through the following inequality:
$$|f_t| \leq \mu f_n,$$
  where $\mu$ is the coefficient of friction at contact. In the advent of slip motion, the equality is satisfied : $|f_t| = \mu f_n$. The value of $\mu$ thus controls the amount of energy dissipated by the flow through frictional contacts. Note that a single coefficient of friction at contact is introduced: we do not distinguish static and dynamical friction.  In addition, a coefficient of restitution $e$ sets the amount of energy dissipated in the advent of a collision, and thus controls the amount of energy dissipated by the flow through collisional contacts. The precise contribution of the two modes of dissipation (collisional and frictional) within a given granular flow is not straightforward to estimate, the multi-contact dynamics particular to dense granular packings being characterized by disorder and complexity \cite{otsuki10}. In the following however, we will not be interested in the particular role of the coefficient of restitution $e$, and we will fix the value of the later to $e =0.5$ coinciding with dense systems.  On the contrary, our interest focuses on the {role of the coefficient of friction $\mu$  on the flow dynamics}.  Hence, its value was alternatively set to $\mu=0.05$ (very small), $\mu =0.5$ (a typical value for  glass beads is $\mu = 0.2$ \cite{foerster94} while $0.5 <\mu < 0.9$ for singing sand grains \cite{dagois10b}) and $\mu =2$ (high). Note that by precluding the use of spring-dashpot models for the contact law, the Contact Dynamics algorithm prevents the introduction of mechanical oscillators in the treatment of grains interactions, and thereby prevent artefact oscillations which may occur in the soft-sphere limit \cite{richard12}. Details on the numerical method can be found in \cite{jean92,radjai96,radjai09}.\\
  The flow configuration investigated  is a 2D periodic flow on erodible bed tilted of an angle $\theta$ (see Figure \ref{Expe}). The grains show a slight size dispersity to avoid ordering effects, too small however to induce segregation: we chose $(d_{max}-d_{min})/d=0.4$, where $d$ is the mean grain diameter ($d=500\mu$m). 
  The packing is obtained by random rain under gravity.  The flow has a periodicity in the longitudinal ($x$) direction; the size $L$ of the simulation cell was set to $45d$ in the greater part of the simulations reported here, corresponding to 3967 grains. However, a specific study of the effect of the value of $L$ on the flow dynamics, and more specifically on the flow oscillatory motion, was performed and  is reported in section \ref{periodicity}. \\
  The erodible bed condition is achieved by trapping grains between vertical walls at the boundary of the simulation cell, thus allowing the upper layer only to flow in response to gravity. The height of the vertical walls is fixed and is $30d$; the height of the unconstrained ({\it ie} free to flow) layer is $h$; depending on the volume fraction ({\it ie} depending on the velocity), $h$ varies between $40d$ and $49d$. The vertical position of the grains $z$  is counted positively following the upward position; the origin is set where the vertical walls stop ({\it ie} at the bottom of the unconstrained layer).  For each value of $\mu$, the tilt angle was varied so as to achieved different flow velocities: $\theta$ ranges from $16^\circ$ to $22^\circ$ for $\mu = 0.05$, from $22^\circ$ to $28^\circ$  for $\mu=0.5$ and from $26^\circ$ to $30^\circ$ for $\mu = 2$. In every cases, stationary regime is reached.  The mean characteristics of the flow thus simulated are detailed in the next section.

\section{Mean velocity and velocity profile}

\begin{figure}
\begin{minipage}{1.\linewidth} 
\centerline{\includegraphics[angle = -90,width = \linewidth]{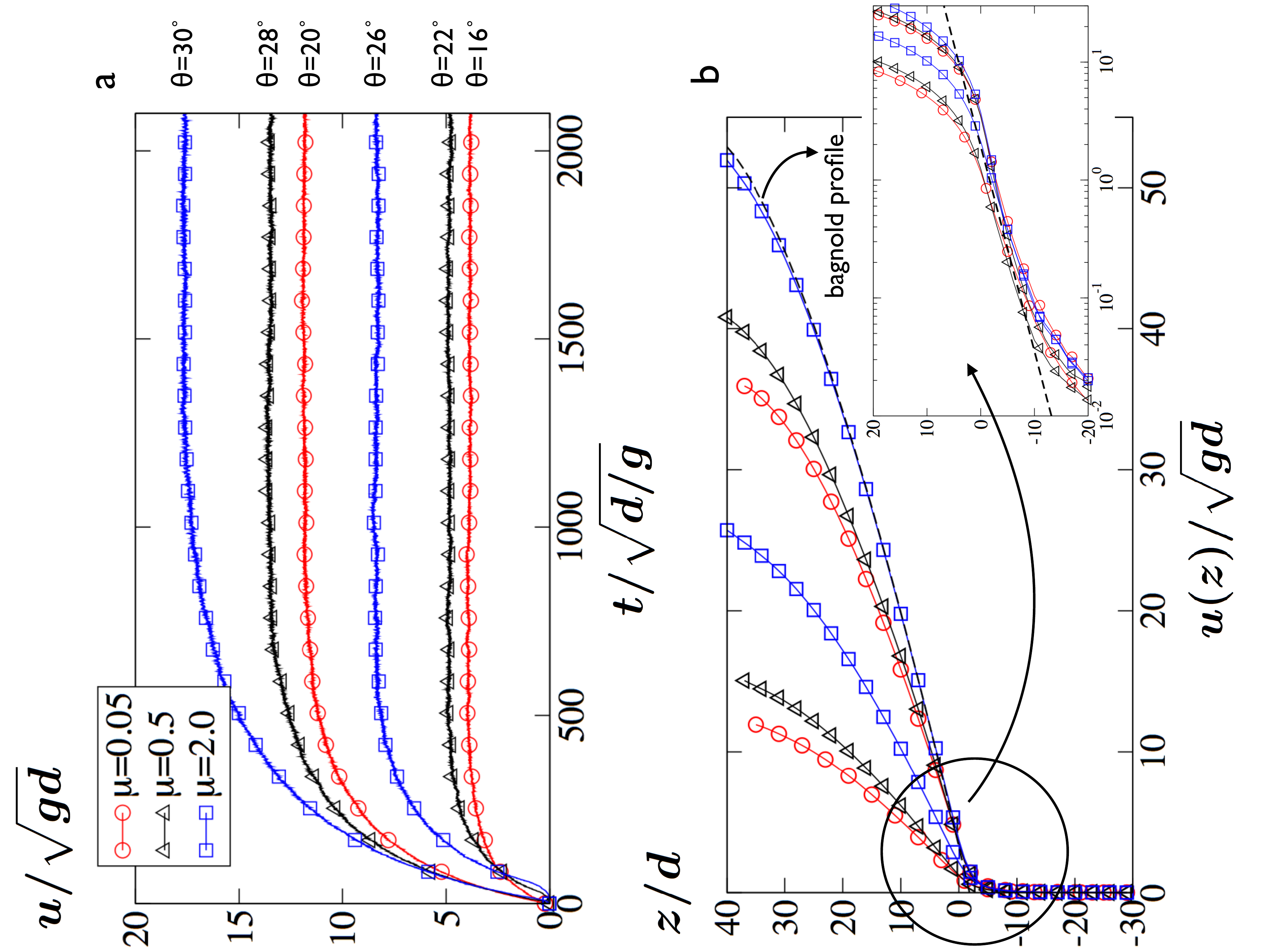}}
\end{minipage}
\caption{{\bf a:} Mean velocity of the grains $u$ (normalized by $\sqrt{gd}$) as a function of time $t$ (normalized by $\sqrt{d/g}$) for different values of the coefficient of friction at contact $\mu=0.05$, $\mu=0.5$ and $\mu=2.0$,  and for different tilt angles $\theta$; {\bf b:} Corresponding velocity profiles time-averaged over the stationary regime; the dotted line shows a Bagnold profile (see equation (\ref{eq:bagnold})); {\bf Inset}: Velocity profiles in semi-log scale; the dotted line shows an exponential decay with a typical length $\lambda=2.5d$. (Color on-line)}
\label{velocity}
\end{figure}

\begin{figure}
\begin{minipage}{1.\linewidth} 
\centerline{\includegraphics[angle =0,width = \linewidth]{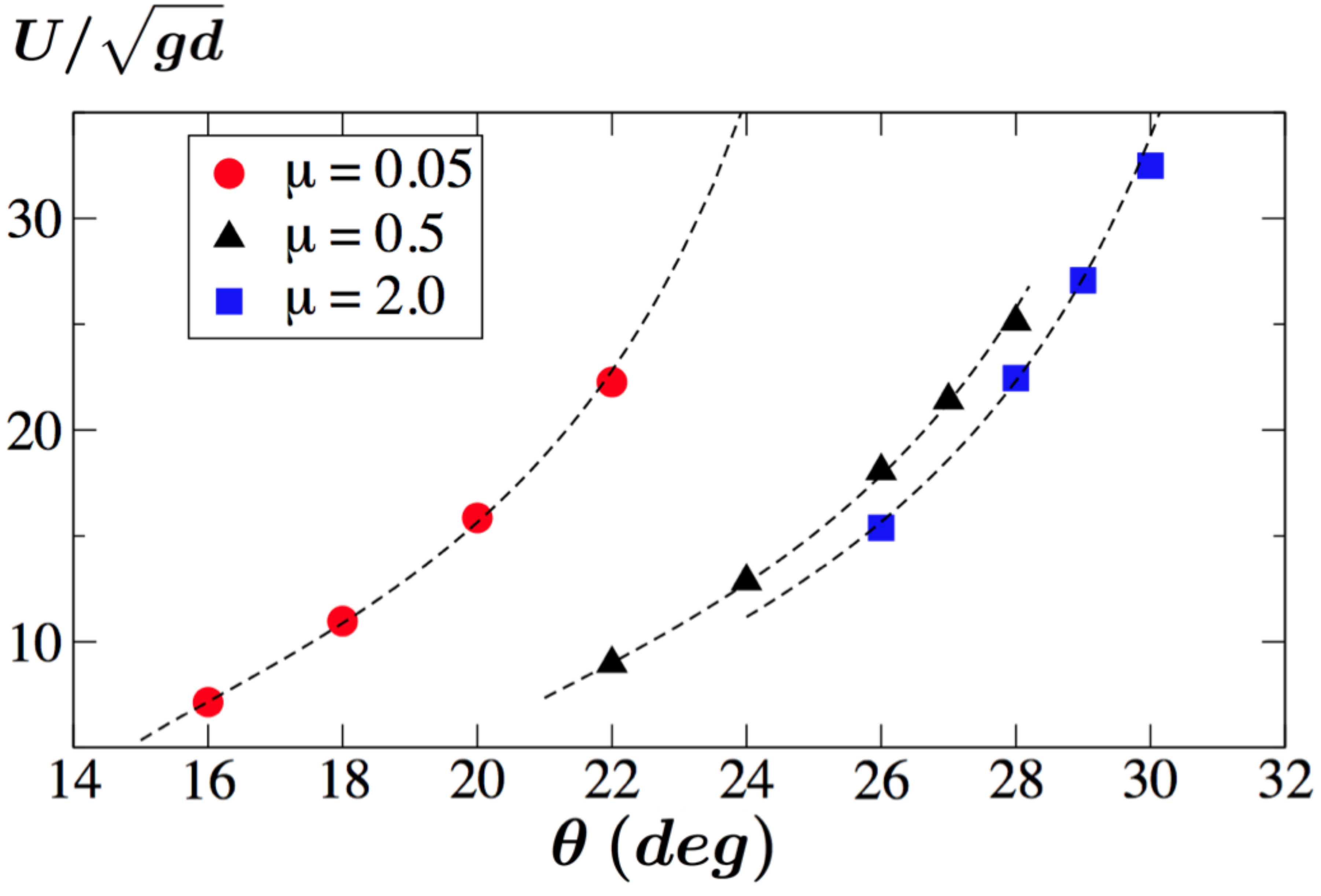}}
\end{minipage}
\caption{Mean flow velocity $U$ of the grains in the flowing layer averaged over the duration of the stationary regime, as a function of $\theta $  for $\mu = 0.05$,  $\mu = 0.5$, and  $\mu = 2$. The dotted lines  shows the experimental relation (\ref{eq:poul}) from \cite{pouliquen99}. (Color on-line) }
\label{Utheta}
\end{figure}

Figure \ref{velocity}-a  shows the time evolution of the mean grain longitudinal (following $x$) velocity $u(t)$ (computed over the total number of grains in the simulation) for the three values of the coefficient of friction at contact $\mu$ (namely 0.005, 0.5 and 2) and for different values of the tilt angle $\theta$. In each case, stationary regime is reached after a transient regime;  we do not study  the latter but focus on the stationary regime in the following. 
For the corresponding runs, Figure \ref{velocity}-b shows the velocity profiles time-averaged over the duration of the stationary regime. We observe well-developed surface flows with a velocity vanishing at the depth corresponding to the upper end of the vertical walls at the boundaries of the simulation cell. 
 We observe that the velocity profiles can be reasonably approximated by a  Bagnold scaling \cite{bagnold66,silbert01}(see Figure \ref{velocity}-b) 
\begin{equation}
u(z) \propto \frac{\sqrt g}{d}\left(h^{3/2} -(h-z)^{3/2} \right), 
\label{eq:bagnold}
\end{equation}
where $h$ is the thickness of the flowing layer, and $z$ is the vertical position of the grains (counted positively in the upward direction).  The Bagnold-like shape of $u(z)$ is in contradiction with experimental and numerical observation of linear profiles for granular flows on erodible beds \cite{komatsu01,bonamy02,courrech05,renouf05,richard08}.  Yet the semi-log plot of the velocity profiles shows the existence of creep motion with an exponential decay  over a typical length $\lambda=2.5 d$, in agreement with experimental findings \cite{komatsu01,bonamy02} (Figure \ref{velocity}-b, inset): we conclude that the erodible bed condition implemented in the numerical simulations reproduces the dissipative properties of a real erodible bed configuration. However, an important difference exists: in real erodible bed conditions, the height of the flowing layer is selected by the flow itself, while in our numerical simulations, the height of the flowing layer is set by the height of the sidewalls. We can show that in our numerical set-up, the shape of the velocity profile is strongly dependent on the relative heights of the flowing layer and the erodible bed. This aspect, although of interest in connection with granular flow rheology, is beyond the scope of the present paper. We will assume that the erodible bed implemented in the simulations, by allowing creep motion to occur, partly reproduce a real erodible flow configuration.\\
Averaging over the duration of the stationary regime, and considering only the grains {in the flowing layer} ({\it ie} grains with a positive z), we compute the mean flow velocity $U$. For the different values of the coefficient of friction at contact $\mu$, $U$ is reported as a function of  the  slope $\theta$  in Figure \ref{Utheta}. The numerical flows obey the chute flow phenomenology observed in \cite{pouliquen99}: 
\begin{equation}
\tan\theta =  \tan\theta_1 + (\tan\theta_2 -\tan\theta_1) \exp \left(-\frac{\beta h}{\ell d}  \frac{\sqrt {gh}}{U}\right),
\label{eq:poul}
\end{equation}
 where $\tan \theta$ identifies with the effective frictional properties of the flow, $\theta_1$  and $\theta_2$ are typical angles dependent on the frictional properties of the grains, $\ell$ is a non dimensional length scale and $\beta=0.136$ (from \cite{pouliquen99}). This approximation is reported in Figure \ref{Utheta} for all values of the friction coefficient at contact $\mu$.  Each value of $\mu$ induces different effective frictional properties of the macroscopic flow. Hence we find $\theta_1=14.0^\circ$ and $\theta_2=28.7^\circ$ for $\mu=0.05$, $\theta_1=18.8^\circ$ and $\theta_2=34.4^\circ$ for $\mu=0.5$, and $\theta_1=19.5^\circ$ and $\theta_2=35.2^\circ$ for $\mu=2.0$; we find $\ell = 2.30$, $2.26$ and $2.29$ respectively. The numerical values corresponding to $\mu=0.5$ are quantitatively consistent with the experimental observation for glass beads \cite{pouliquen99}. \\ 
 Alternatively, we can show that the flow satisfy the $\mu(I)$ dependence by defining $I= dU/(\sqrt{g}h^{3/2})$ \cite{gdrmidi04,dacruz05}: 
 \begin{equation}
 \tan\theta = \tan \theta_1 + \frac{\tan\theta_2-\tan\theta_1}{I_0/I +1}. 
 \label{eq:muI}
 \end{equation}
 The best fist gives $I_0 = 0.33$, and  $\theta_1=11.03^\circ$ and $\theta_2=33.02^\circ$ for $\mu=0.05$, $\theta_1=15.38^\circ$ and $\theta_2=38.83^\circ$ for $\mu=0.5$, and $\theta_1=16.17^\circ$ and $\theta_2=39.35^\circ$ for $\mu=2.0$.\\
 The analysis of the mean velocity and velocity profiles of the numerical flows thus show that the latter behave accordingly to experimental evidence \cite{pouliquen99,gdrmidi04}.

\begin{figure}
\begin{minipage}{1.\linewidth} 
\centerline{\includegraphics[angle = -90,width = \linewidth]{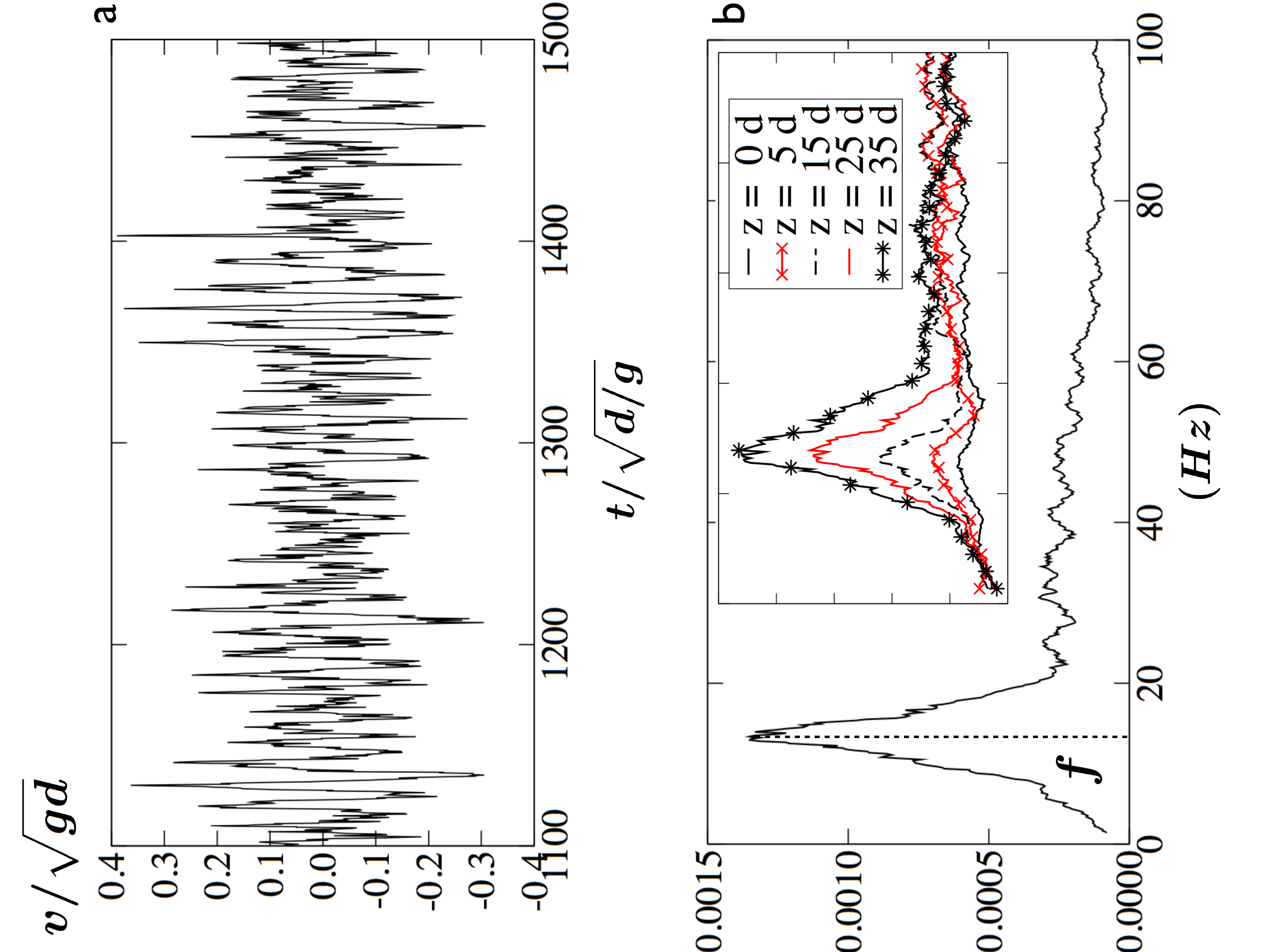}}
\end{minipage}
\caption{{\bf a:} Instantaneous normal velocity $v$ (normalized by $\sqrt{gd}$) as a function of time $t$ (normalized by $\sqrt{d/g}$); {\bf b:} Fourier transform of $v(t)$ over the duration of the stationary regime showing a peak frequency $f$;  {\bf Inset:} Fourier transform of the instantaneous normal velocity at different depth in the flow.  (Color on-line)}
\label{fvt}
\end{figure}

\section{Oscillatory motion: frequency and amplitude}

\begin{figure}
\begin{minipage}{1.\linewidth} 
\centerline{\includegraphics[angle = 0,width = \linewidth]{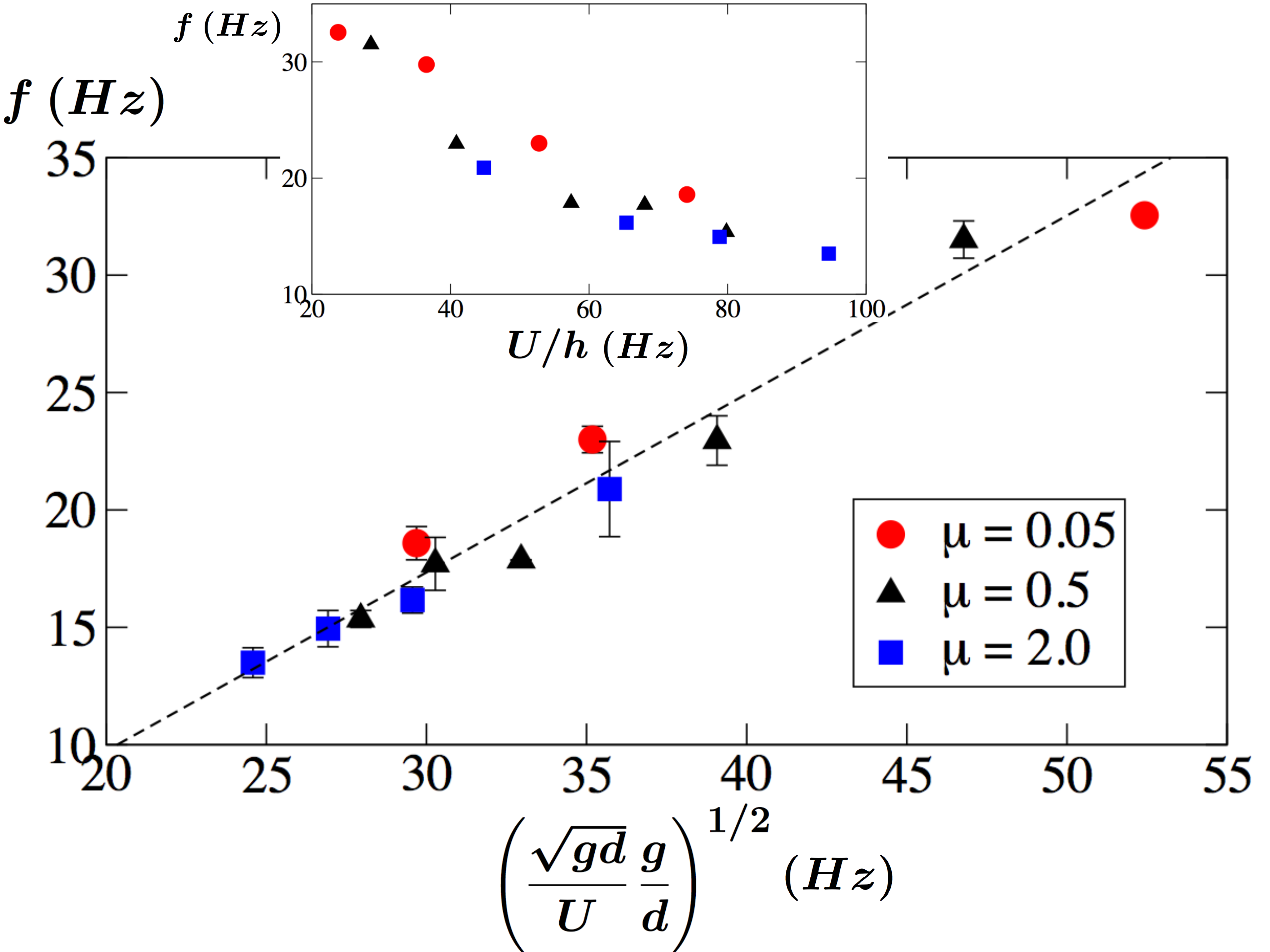}}
\end{minipage}
\caption{Peak frequency $f$ (Hz) of the Fourier transform of the instantaneous normal flow velocity $v(t)$ as a function of $\left( \frac{\sqrt{gd}}{U} \times \frac{g}{d} \right)^{1/2}$ (Hz). {\bf Inset}:  Peak frequency $f$ (Hz) of the Fourier transform of the instantaneous normal flow velocity $v(t)$ as a function of the shear rate $U/h$ (Hz). (Color on-line)}
\label{fU}
\end{figure}

\begin{figure}
\begin{minipage}{1.\linewidth} 
\centerline{\includegraphics[angle = -90,width = \linewidth]{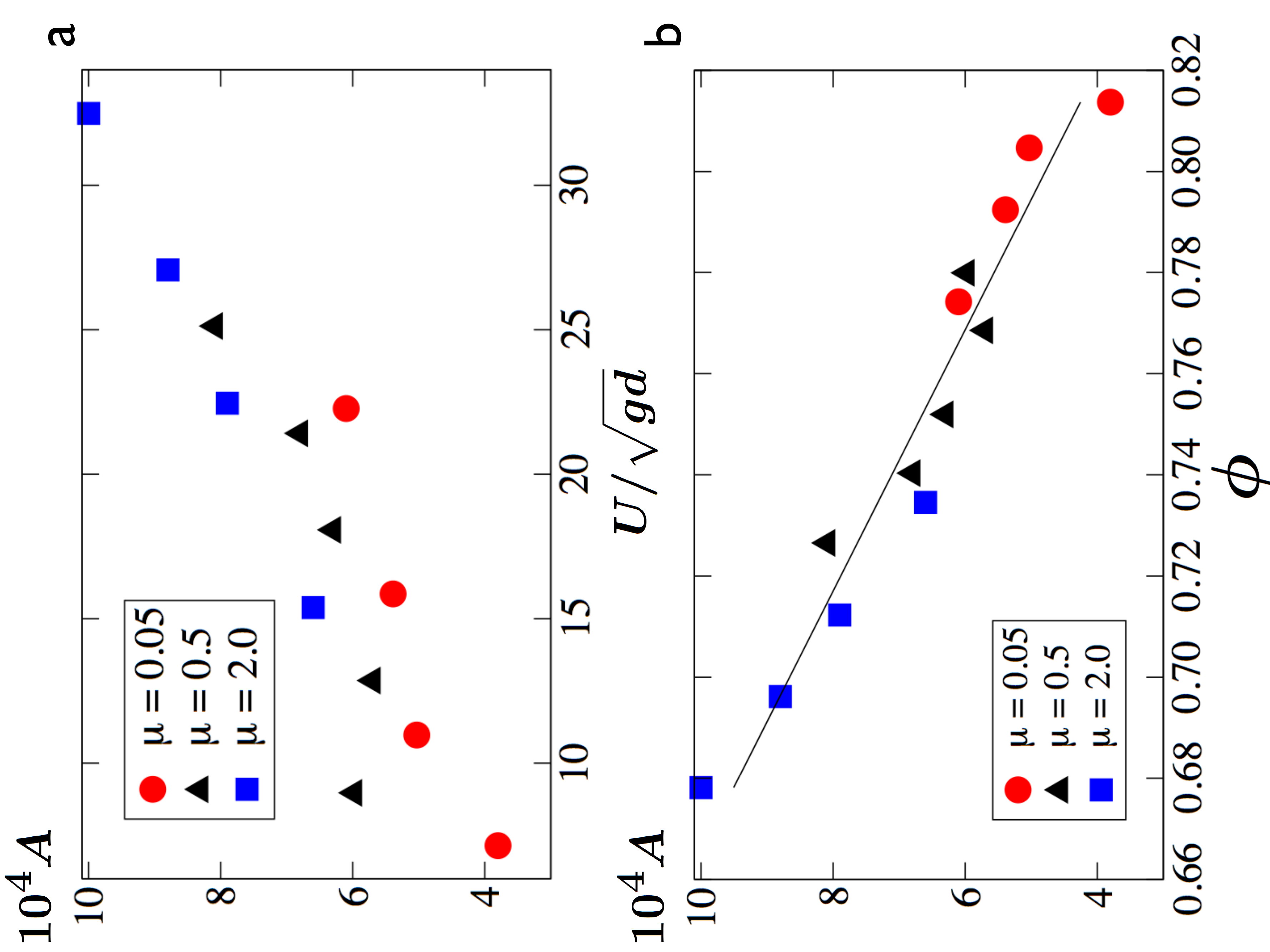}}
\end{minipage}
\caption{{\bf a:} Amplitude $A$ ($\times 10^4$) of the frequency peak as a function of the flow mean longitudinal velocity $U$ (normalized $\sqrt{gd}$).  {\bf b:} amplitude $A$ ($\times 10^4$) of the frequency peak as a function of the flow volume fraction $\phi$. (Color on-line) }
\label{AUPhi}
\end{figure}

The mean normal velocity V of the grains ({\it ie} in the z- direction), averaged over time, is expectedly zero, or the flow would expand infinitely and eventually be turned into a dilute gas.  However, the instantaneous normal velocity $v$, plotted as a function of time, shows clear rapid oscillations around zero, revealing an oscillatory motion implying successions of dilation phase and compaction phase. Figure \ref{fvt}-a shows, as an example, the time variation of the instantaneous normal velocity of the grains $v(t)$ over a short time interval for $\mu =2.0$ and $\theta =30^\circ$.  To establish whether this oscillatory dynamics involves a characteristic frequency, we compute the Fourier transform of $v(t)$ over the duration of the stationary regime (Figure \ref{fvt}-b): a well-defined peak frequency emerges;  we denote this peak frequency $f$ in the following. The inset in Figure \ref{fvt}-b shows the Fourier transform of the instantaneous normal velocity at different depth in the flowing layer. We observe that the amplitude of the peak frequency rapidly decreases  and eventually vanishes as we go deeper in the flow, whereby we conclude that the oscillatory motion is a surface phenomenon rather than a bulk one. \\
For all simulations, we measure the peak frequency $f$  characterizing the oscillation over the stationary regime (error bars are evaluated systematically based on the width of the frequency peak).  We observe  values ranging from $10$ to $40$ Hz, that is significantly smaller than the typical frequency attached to the grain size $\sqrt{g/d} =140$ Hz.  The values observed for $f$ are also below the values measured for sonic sands, typically of $70$ to $110$ Hz \cite{hunt10}. A reasonable correlation suggests that  the value of $f$ increases like $\left( \frac{\sqrt{gd}}{U} \times \frac{g}{d} \right)^{1/2} $, although the range of values explored does not allow for a discussion on the shape of the dependence  (Figure \ref{fU}): hence we will not address the possible origin of this correlation in the following. Interestingly,  the frequency characterizing the oscillation does not identify with the shear rate: Figure \ref{fU}  shows an inverse correlation between $f$ and $U/h$.  \\
The trend relating frequency and velocity displayed in Figure \ref{fU} is not dependent on the  value of the coefficient of friction $\mu$: the latter seems to be  playing no particular role in the value of $f$.
However, oscillatory motion is not characterized only by  the peak frequency $f$, but also by the amplitude $A$ of the peak. Figure \ref{AUPhi}-a  shows  the amplitude $A$ ($\times 10^4$) as a function of the flow mean longitudinal velocity $U$:  we observe a positive correlation, with larger values of $\mu$ inducing larger amplitudes.  High velocities and high friction at contact thus induce larger oscillations. This double influence can be summed-up when plotting $A$ as a function of the flow mean volume fraction $\phi =$ volume-of-grains/total-volume (computed in the flowing layer over the duration of the stationary regime) (Figure \ref{AUPhi}-b): the data collapse and  show that larger amplitudes coincide with dilute flows, while denser flows induce a smaller amplitude. 


\section{On the role of friction}

\begin{figure}
\begin{minipage}{1.\linewidth} 
\centerline{\includegraphics[angle = 0,width = \linewidth]{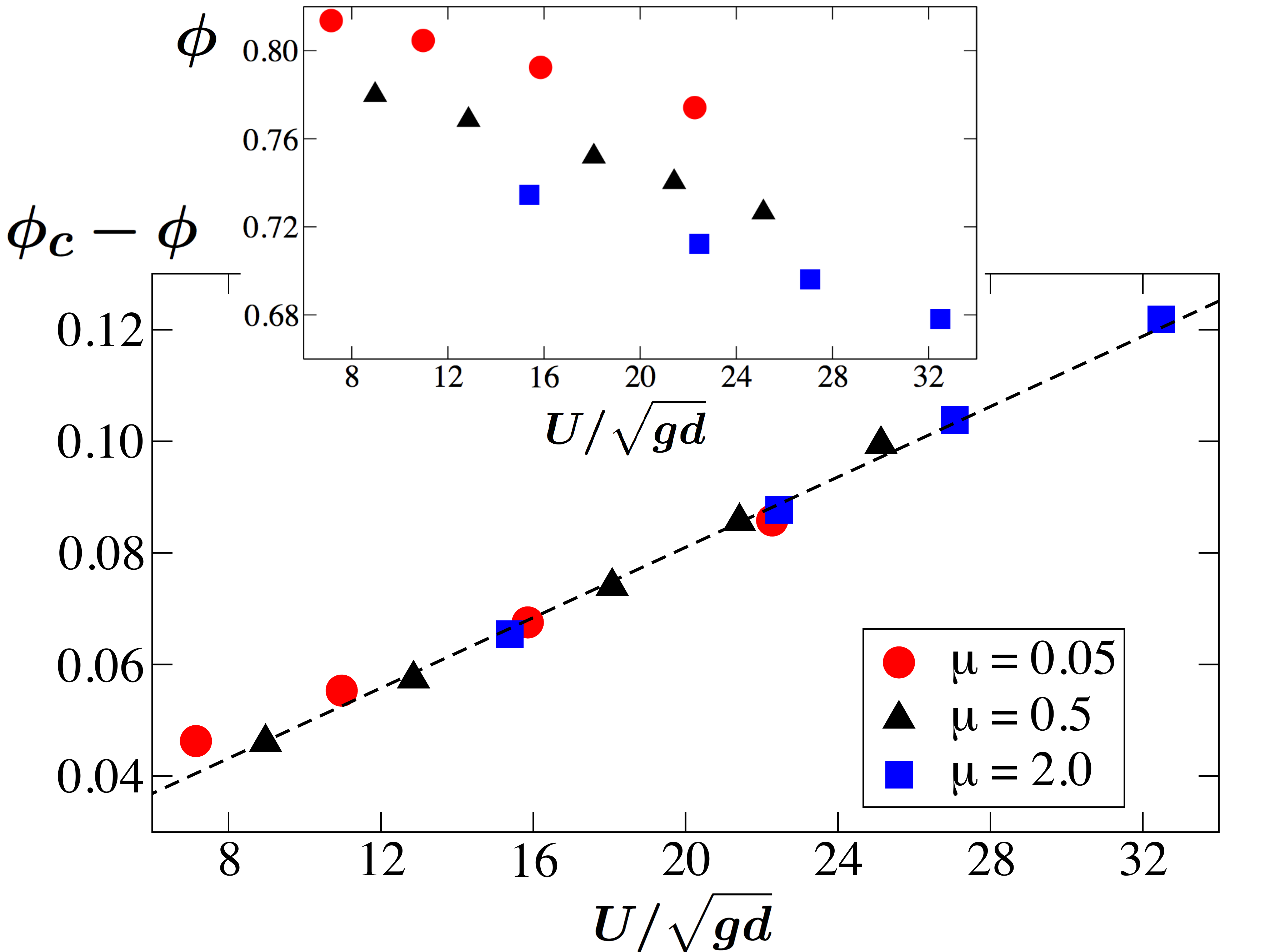}}
\end{minipage}
\caption{{\bf Inset}: Flow mean volume fraction $\phi$ as a function of the grains mean velocity $U$ (normalized $\sqrt{gd}$) for the different values of the coefficient of friction at contact $\mu$. {\bf Main graph}: $\phi-\phi_c$ as a function of $U$ (normalized $\sqrt{gd}$), where $\phi_c$ is a constant whose value depends on $\mu$ (see equation (\ref{eq:phi})); the dotted line shows a linear fit with slope $0.00315$. (Color on-line) }
\label{PhiU}
\end{figure}

The volume fraction $\phi$ of a granular flow is dependent on the dynamics: rapid flows coincide with dilute states, {\it ie} smaller volume fractions \cite{gdrmidi04,dacruz05,staron10}. The relation between the two involves the frictional properties of the material: as shown in Figure \ref{PhiU} (inset), plotting $\phi$ as a function of $U$ reveals three distinct series of points coinciding with each value of $\mu$. This behavior can be described by the following dependence: 
\begin{equation}
 \phi = \phi_c(\mu) - k \frac{U}{\sqrt{gd}},
 \label{eq:phi}
 \end{equation}
where $k= 0.00315 $ and $\phi_c$ is a decreasing function of $\mu$: $\phi_c = 0.86$ for $\mu = 0.05$, $\phi_c=0.826$ for $\mu = 0.5$ and $\phi_c=0.800$ for $\mu=2$ (see Figure \ref{PhiU}). In other words, for a given velocity $U$, higher contact friction implies smaller volume fraction. The dependence of $\phi$ on the dynamics was already reported in \cite{gdrmidi04,dacruz05} in terms of the inertial number $I=dU/(\sqrt{g}h^{3/2})$: $\phi = \phi_M +(\phi_m -\phi_M) I $, where  $\phi_M$ and $\phi_m$ are extremal values of the volume fraction.\\
 Relation (\ref{eq:phi}) thus gives a possible explanation for the role of contact friction in the oscillatory motion of granular flows: higher friction at contact between the grains is responsible for smaller volume fraction, thus leading to larger amplitude of grains vertical motion.  Translated in a natural case, larger friction (as induced by the mineral coating observed at the surface of sonic sands) may lead to a larger amplitude of the oscillatory motion, and possibly due to that, to the emission of an audible sound.  In the same way, relation (\ref{eq:phi}) renders the fact that larger velocities induce more dilute flows, thus  larger amplitude of grains vertical motion, which may explain why a minimum flow velocity is necessary for sonic sands to emit sound \cite{andreotti04,douady06,dagois10,andreotti12}.

\section{Influence of the size of the simulation cell}
\label{periodicity}

\begin{figure}
\begin{minipage}{1.\linewidth} 
\centerline{\includegraphics[angle = -90,width = \linewidth]{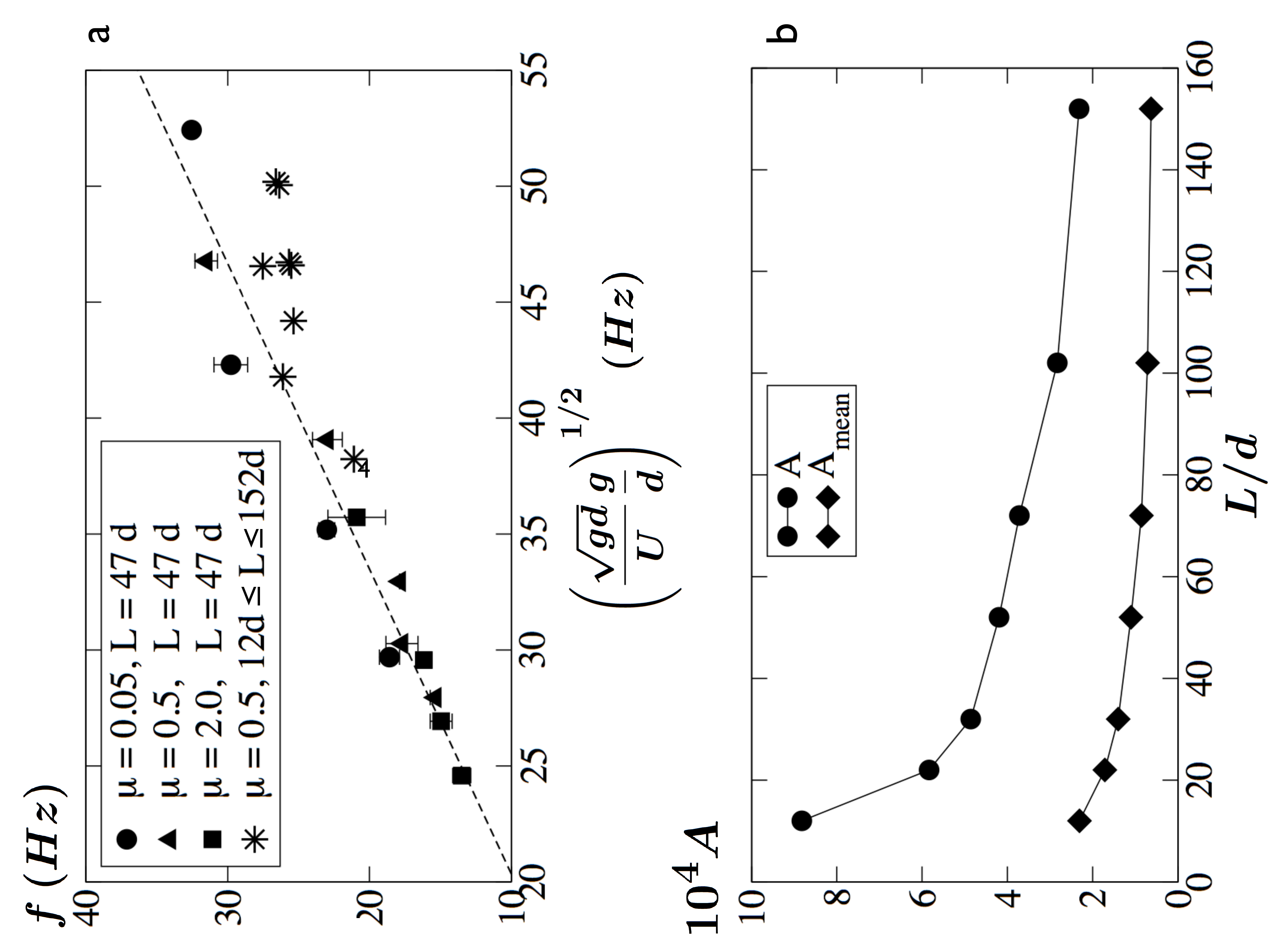}}
\end{minipage}
\caption{ {\bf a:} Peak frequency $f$ of the Fourier transform of the instantaneous normal flow velocity $v(t)$ as a function of  $\left( \frac{\sqrt{gd}}{U} \times \frac{g}{d} \right)^{1/2}$. {\bf b:} Amplitude $A_\text{mean}$ ($\times 10^4$) of the Fourier transform of the normal velocity $v(t)$ at high frequencies  and the amplitude $A$ ($\times 10^4$) of the peak frequency $f$.}
\label{AfUL}
\end{figure}

The existence of a spatial periodicity in the numerical systems studied here may be suspected to affect the time periodicity  characterizing the flow dynamics.  The length of the simulation cell being $L$, and the mean flow longitudinal velocity being $U$, the obvious time scale related to the geometrical periodicity of the system is $L/U$. If the geometrical periodicity was to affect the time periodicity, then we would expect the latter  to exhibit a frequency scaling like $U/L$. This is very different from the frequency $f$  emerging from the analysis of the oscillatory motion of the numerical flows, and the associated dependence on the velocity $U$ (Figure \ref{fU}) : $f$ increases with $\left( \frac{\sqrt{gd}}{U} \times \frac{g}{d} \right)^{1/2}$. Hence, it seems unlikely that the periodicity of the simulation cell has a significant influence on the oscillatory motion observed. This however needs clarification. \\
Therefore, we perform series of simulations where $L$ is varied: $L=12d$, $22d$, $32d$, $52d$, $72d$,  $102d$,  and $L=152d$.  The coefficient of friction is fixed: $\mu=0.5$, as well as the slope $ \theta=22^\circ$.  As previously, we analyze the mean velocity of the flow  and the oscillatory motion visible in the time fluctuations of the normal velocity $v(t)$.  The peak frequency $f$ is reported as a function of $\left( \frac{\sqrt{gd}}{U} \times \frac{g}{d} \right)^{1/2}$ in Figure  \ref{AfUL}-a: although scattering occurs,  the value of $L$ is not significantly affecting the dependence  already  observed in Figure \ref{fU}. \\
Yet, the size of the simulation cell $L$ might play a role in the amplitude of the oscillatory motion, and possibly influence its very existence.  We denote $A_\text{mean}$ the amplitude of the Fourier transform of $v(t)$ at high frequencies ({\it ie}  $100$Hz $\leq f $). The value of  $A_\text{mean}$ is reported is Figure \ref{AfUL}-b as a function of $L$: we observe indeed that smaller systems favor larger oscillations. However, the amplitude saturates towards a minimum value for larger systems, showing that the oscillation is not the result of the system size.  Accordingly, we expect the amplitude $A$ of the peak frequency $f$ to decrease and saturate for larger value of $L$.  This is indeed what is observed  (Figure \ref{AfUL}-b). \\
There is however an aspect in the present  analysis which prevents us from interpreting further the amplitude $A$ of the oscillation: the spatial localization of the latter. Indeed, so far we have processed the  normal velocity $v(t)$ averaged over the {\em whole} flowing layer. If the oscillation is localized in space,  part of the information is "watered down" by including the grains dynamics of the whole flow. This is all the more likely to happen that the system is large. It is thus probable that Figure \ref{AfUL}-b shows not only the effect of the periodicity, but also the information loss due to averaging over the system size. We can conclude nevertheless that the system size $L$ plays no crucial role in the results discussed above.

\section{Conclusion}
This contribution reports on numerical simulations of 2D granular flows on erodible beds. The broad aim of this work is to investigate {\it i)}  whether  simple flows of model granular matter exhibits spontaneous oscillatory motion in generic flow conditions, and in this case, {\it ii)} whether the frictional properties of the contacts between grains may affect the existence or the characteristics of this oscillatory motion. 
The analysis of different series of simulations show that the flow develops an oscillatory motion with a well-defined frequency which increases like the inverse of the velocity's square root. We show that the oscillation is essentially a surface phenomena. The amplitude of the oscillation is higher for lower volume fractions. It can thus be related to the flow velocity, higher velocities favoring lower volume fraction. For the same reason, it is also dependent on grains friction properties: indeed, large contact friction is found to induce lower volume fraction, and thus larger amplitude.  The study of the influence of the periodic geometry of the simulation cell shows no  significant effect. \\
Although  one should be careful while drawing analogy between simple numerical models and nature, it is interesting to discuss these results in relation to sonic sands. Indeed, this work suggests that surface oscillation is likely to develop during the flow of granular matter, and that its amplitude is dependent on both velocity and grains properties, in agreement with observation. How this oscillation develops into a loud audible sound is beyond the reach of simple 2D discrete simulations; however they provide an interesting insight into basic mechanisms that may be relevant to the complex question of how sand may produce sound.


%
%
%

\end{document}